\pdfoutput=1
\documentclass{aa}  

\usepackage{graphicx}
\usepackage{txfonts}
\usepackage{natbib}
\bibpunct{(}{)}{;}{a}{}{,}
\def\aap{A\&A}
\def\apj{ApJ}
\def\apjs{ApJS}
\def\aj{AJ}

\def \hi {\ion{H}{i}}
\def \ovi {\ion{O}{vi}}

\newcommand {\kms} {\,{\rm km\,s}^{-1}}

\newcommand {\pc} {\,{\rm pc}}

\newcommand {\kpc} {\,{\rm kpc}}

\newcommand {\cmmq}{\,{\rm cm^{-2}}}

\newcommand {\kmskpc} {\,{\rm km\,s}^{-1}\,{\rm \kpc}^{-1}}

\newcommand {\de}{^{\circ}}

\newcommand {\msun}{\,{M}_\odot}

\newcommand{\K}{\,{\rm K}}
\newcommand{\mK}{\,{\rm mK}}

\newcommand {\vlos}{$v_{\rm LOS}$\ }
\newcommand {\vdev}{$v_{\rm DEV}$\ }

\def\deg{\hbox{$^\circ$}}

\begin{document}
   \title{Modelling the \hi\ halo of the Milky Way}

   \author{A. Marasco \& F. Fraternali}

   \institute{Astronomy Department, University of Bologna,
via Ranzani 1, I-40127, Bologna, Italy\\ \email{antonino.marasco2@unibo.it; filippo.fraternali@unibo.it}}
      
  \titlerunning{Modelling the \hi\ halo of the Milky Way}

   \date{Received / Accepted}

  \abstract 
  {}
  {We studied the global distribution and kinematics of the extra-planar  
neutral gas in the Milky Way.}
  {We built 3D models for a series of Galactic \hi\ layers, projected them  
for an inside view, and compared them with the Leiden-Argentina-Bonn 21-cm observations.}
 {We show that the Milky Way disk is surrounded by an extended halo of neutral gas with a vertical scale-height of $1.6^{+0.6}_{-0.4}\kpc$ and an \hi\ mass of $3.2^{+1.0}_{-0.9}\times10^8 M_{\odot}$, which is $\sim5\!-\!10\%$ of the total Galactic \hi.
This \hi\ halo rotates more slowly than the disk with a vertical velocity gradient of $-15\pm4\kmskpc$.
We found evidence for a global infall motion in the halo, both vertical ($20^{+5}_{-7}\kms$) and radial ($30^{+7}_{-5}\kms$).}
{The Milky Way \hi\ extra-planar layer shows properties similar to the halos of external galaxies and it is compatible with being predominantly produced by supernova explosions in the disk.
It is most likely composed of distinct gas complexes with masses of $\sim 10^{4-5}M_{\odot}$ of which the Intermediate Velocity Clouds are the local manifestations.
The classical High Velocity Clouds appear to be a separate population.
}
  \maketitle
%

\section{Introduction}\label{intro}

In the recent years it has become clear that neutral hydrogen in nearby galaxies is not confined to a thin disk, but it extends for several kiloparsecs into the halo region. Several processes may contribute to create this \emph{halo} (or \emph{extra-planar}) gas: galactic fountains and extra-galactic accretion are the main candidates. 
In the first case the gaseous halo is built up by the supernova feedback. 
In the classical picture \citep{Shapiro1976,Bregman1980} supernovae explosions heat the gas in the disk and eject it up to a distance of few kiloparsecs, the gas travels through the halo region, it cools and eventually falls back to the plane.
In the latter scheme the halo originates from gas accretion out of the intergalactic environment \citep{Oort1970,Kaufmann2006}. 
The chemical and kinematical properties of the halo gas expected in this second case can differ significantly from the properties of the gas in the Galactic disk.

Observations of external galaxies suggest that both processes contribute. A large amount of halo gas has probably an internal origin. High velocity outflows of neutral and ionized gas from the disk are observed to be correlated with star forming regions \citep{Kamphuis1991,Fraternali2004}. 
The extra-planar neutral hydrogen has a kinematics similar to that of the disk \hi\  \citep{Fraternali2002} but with some peculiar properties: a) it rotates more slowly \citep{Oosterloo2007}; b) it shows vertical motions \citep{Boomsma2005}; c) it has a global inflow \citep{Fraternali2001}. Extra-planar diffuse ionized gas (eDIG) is present in all galaxies with sufficiently high star formation rates \citep{Rossa2003} and its kinematics seems to follow closely that of the neutral component \citep{Heald2006,Kamphuis2007}.

A fraction of the halo gas may also have an external origin: some galaxies (including the Milky Way) are surrounded by clouds or filaments of \hi\ with highly anomalous kinematics, as in the cases of M\,31 \citep{Thilker2004} and NGC\,2403 \citep{Fraternali2002}, that may be the signature of gas accretion and minor mergers.
This extra-galactic gas accretion is of vital importance to sustain the star formation in galaxies like the Milky Way \citep{Sancisi2008}. 

In the Milky Way the Intermediate and High Velocity Clouds \citep[IVCs and HVCs,][]{Wakker1997} constitute the main evidence for the presence of extra-planar gas. 
The HVCs are clouds of neutral hydrogen with a line-of-sight velocity significantly anomalous with respect to the Galactic rotation. 
They have been known for half a century \citep{Munch1961} but only in the last decade observations have unequivocally confirmed that they belong to the halo region \citep{Wakker2004,Wakker2008} at distances from the Sun $\lesssim\!10\kpc$. 
Also, they are observed to have low metallicity ($\sim\!0.1\,Z_{\odot}$), which suggests an external origins for most of these objects \citep[see for instance][for complex C]{Tripp2003}. 
In contrast, the IVCs (\hi\ clouds with a lower anomalous velocities) seem to constitute a different population, since their distances are lower ($\sim1\kpc$), their metallicity is higher ($\sim1\,Z_{\odot}$) and their kinematics follow more closely the Galactic rotation. 
Therefore these clouds have been regarded as possible galactic fountain objects \citep{Wakker2001}.

Anomalous \hi\ clouds are not the only observable component of the Galactic halo. 
A large amount of diffuse \hi\ emission has been found by \citet{Lockman1984} in the lower halo regions of the inner Galaxy; \citet{Levine2008} analysed this gas within $0.1\kpc$ of the mid-plane deriving a vertical lag in the rotation velocity of $-22\kmskpc$. 
Signs of a local halo building from recent stellar feedback are visible in the gigantic Ophiucus superbubble, this is a huge structure of neutral and ionized hydrogen reaching heights of $\sim\!3.4\kpc$ above the disk \citep{Pidopryhora2007}. 
The presence of an extra-planar DIG layer in the Galaxy has been known since long time \citep{Reynolds1973}.
Measurements of \ovi\ absorption lines in the Galactic halo obtained with the Far Ultraviolet Spectroscopic Explorer \citep{Sembach2006} confirmed the presence of a `hot Galactic halo' extended over tens of kiloparsecs, already hypothesized by \citet{Spitzer1956} as a medium to pressure-confine the HVCs.

In the literature there is little attempt to describe the \hi\ halo of the Milky Way as a global structure. 
\citet[][hereafter KD08]{KD08} described the whole \hi\ distribution in the Galaxy via a spatial reconstruction of the Leiden-Argentine-Bonn \citep[LAB,][]{Kalberla2005} \hi\ Survey dataset. 
The de-projection of the data relies sensitively on the rotation curve assumed, which has been derived by modelling the Galactic dark matter halo \citep{Kalberla2007} and under the assumption that gravity dominates the dynamics of the gas everywhere. 
KD08 found that $10\%$ of the Galactic \hi\ is located outside the disk and it is highly turbulent.

In this paper we model the global \hi\ component that surrounds the disk of our Galaxy. 
We make no assumption about the dynamical state of this gas but simply derive its global properties directly from the data. 
Section \ref{modellingthehalo} describes how we modelled the halo emission and what is the effect of the various kinematical parameters. In Section \ref{gradient} we derive the vertical rotational gradient in the inner halo and in Section \ref{results} we estimate the kinematical parameter of the Galactic halo. A discussion of the physical interpretation of these parameters follows in Section \ref{discussion}.

\section{Modelling the \hi\ halo}\label{modellingthehalo}

We assumed that the Galactic \hi\ emission is due to two distinct components: the disk and the halo. 
The physical and kinematical properties of the first are assumed to be known (Section \ref{disk}). 
In contrast, the halo properties are not known and its distribution and kinematics have to be derived by fitting a number of free parameters.
We modelled the Galactic \hi\ emission by building pseudo-datacubes with the same resolution and total flux as the LAB dataset, and we compared them with the latter.

In the following we will use two coordinate systems. The first is the Galactic coordinate system, where $l$ is the Galactic longitude, $b$ the Galactic latitude, and \vlos is the velocity along the line of sight. The second is a cylindrical system centred at the Galactic Centre, where $R$ is galactocentric radius, $\phi$ the azimuthal angle (set to $0\de$ at $l=0\de$) and $z$ is the height from the mid-plane. 

\subsection{Removing the \hi\ disk emission} \label{disk}

In order to better visualize the extra-planar \hi\ emission in the Milky Way we first removed the emission associated with the disk applying the standard technique that makes use of the \emph{deviation velocity} \citep{Wakker1991}.
For a fixed $(l,b)$ pair, the deviation velocity $v_{\rm DEV}$ is the difference between the observed \vlos and the largest velocity predicted by a disk model. 
It depends on the rotation curve and on the disk geometry adopted.
We assumed circular orbits and a flat rotation curve, independent of $z$, with rotation velocity $v_{\phi}(R,z) = v_{\odot} = 220\kms$. 
For the disk geometry we used the flared disk described in KD08, with the scale-height given by $h_{\rm s}(R) = h_{\odot}e^{(R-R_{\odot})/R_{\rm f}}$, where $h_{\odot}=0.15\kpc$ and $R_{\rm f}=9.8\kpc$. 
The $h_{\rm s}(R)$ holds for $R>5\kpc$ and it is constant for inner radii.

The line-of-sight velocity of a cloud at a distance $d$ from the Sun in a generic direction $(l,b)$ is given by:
\begin{equation}\label{vlos}
v_{\rm LOS}(l,b,d) = \left(v_{\phi}(R,z)\frac{R_{\odot}}{R}-v_{\odot}\right)\sin(l)\cos(b)
\end{equation}
where 
$R_{\odot}=8.5\kpc$ is the distance of the Sun from the Galactic Centre. 
The maximum and the minimum values of eq.\ (\ref{vlos}) (the `envelope' of $v_{\rm LOS}$) have to be determined by varying $d$ in the range $[0, d_{\max}]$, where $d_{\max}$ is $d(R_{\max}, z_{\max})$. 
We assumed $R_{\max}=35\kpc$ (as in KD08) and $z_{\max}$ is chosen to be $7\times h_{\rm s}(R)$, i.e. about $1\kpc$ in the central parts.
With these assumptions the envelope is large, ensuring that the disk emission is completely removed.
Given this envelope, \vdev can be easily estimated for any $(l, b)$; in agreement with the literature \citep{Wakker2004}, we cut out data with $|v_{\rm DEV}|<35\kms$.
We also removed the \hi\ emission of the Outer Arm region \citep[a distant and bright spiral arm that belongs to the warp;][] {Habing1966}, the Magellanic Clouds, the Magellanic Stream and external galaxies. 
We stress that the geometrical parameters used to describe the shape of the disk weakly affect the deviation velocity calculation.
For example, assuming a disk with no flare and $2\kpc$ of thickness gives an envelope of $v_{\rm LOS}$ similar to one above. 

\begin{figure*}[hbt!]
\centering
\includegraphics[width=0.75\textwidth]{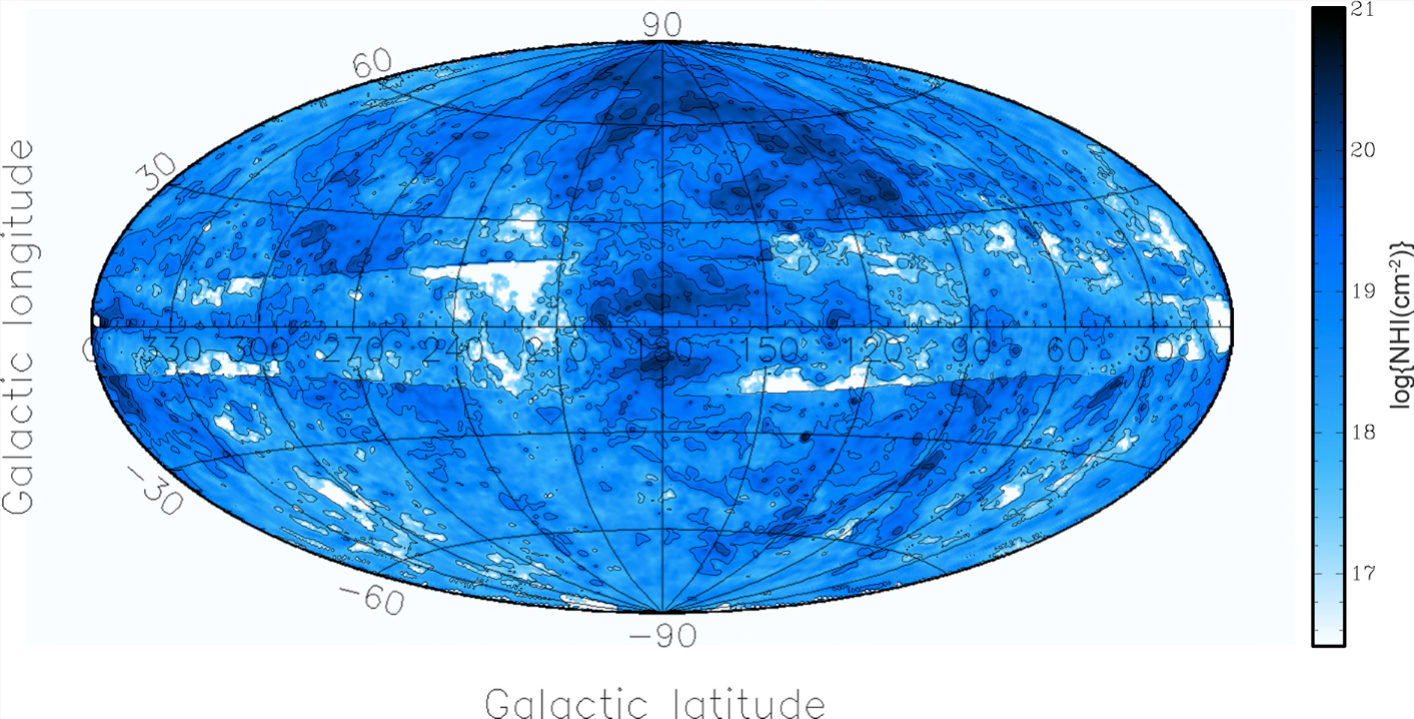}
\caption{All-sky map of the \hi\ halo of the Milky Way:
velocity integrated \hi\ map obtained from the LAB Survey after removing the disk emission, the Magellanic Clouds and Stream and the Outer Arm.
The sharp edges close to the mid-plane are artifacts of the disk removal.
The dataset has been smoothed to $1\de$ resolution. 
Contour levels are: $0.5$, $10$, $50$ and $120\times10^{18}\cmmq$ in column density.}
\label{allskyhalo}
\end{figure*}

Figure \ref{allskyhalo} shows the resulting velocity-integrated \hi\ emission obtained after the disk exclusion.
All the \hi\ emission in Fig.\ \ref{allskyhalo} is not compatible with a thin disk in differential rotation.
Some HVCs and IVCs can be recognized in this map, and they appear embedded in a more diffuse extended medium. 
This medium is visible in almost any direction of the sky and it is the main subject of our investigation.

\subsection{The \hi\ halo distribution and kinematics}\label{halo}

We modelled the \hi\ halo of the Milky Way assuming a density distribution $\rho(R,z)$ analogous to that used by \citet{Oosterloo2007} for the nearby galaxy NGC\,891, described by the formula:

\begin{equation}\label{halo891}
\rho(R,z) = \rho_0 \left(1+\frac{R}{R_{\rm g}}\right)^{\gamma}\exp\left(-\frac{R}{R_{\rm g}}\right){\rm{sech}}^2\left(\frac{z}{h_{\rm g}} \right)
\end{equation}

where $R_{\rm g}=1.61\kpc$ and $\gamma=4.87$, as recently estimated by \citet{Marinacci2010b}. For a given kinematics, the only parameter that critically influences the resulting pseudo-datacube is the scale-height, while the scale-radius $R_{\rm g}$ has a minor effect. Therefore we set $h_g$ as a free parameter, and we fixed $R_{\rm g}$ and $\gamma$ to the above values. Also, we set the central density $\rho_0$ so that the total flux of the model equals the flux of the LAB data, once the disk emission has been excluded.

We modelled the kinematics of the halo using three parameters: 
\begin{enumerate}
\item the vertical gradient of rotation velocity $dv_{\phi}/d|z|$, so that $v_{\phi}(z)=v_{\odot}+|z|\times dv_{\phi}/d|z|$ for every $R$; 
\item the velocity along the z-direction $v_z$, assumed positive if the gas is escaping;
\item the velocity along the R-direction $v_R$, assumed positive if the gas is moving to larger radii.
\end{enumerate}
At first, we assumed that these parameters do not depend on $R$ and $z$. The resulting velocity field ${\bf{v}}=(v_R, v_{\phi}, v_z)$ has to be projected along the line of sight to obtain \vlos:
{\setlength\arraycolsep{-28pt}
\begin{eqnarray}
v_{\rm LOS}(l,b,d) = v_R\left(\frac{R^2+d^2-R_{\odot}^2-z^2}{2\,R\,d}\right)+v_z\sin(b)+{}\nonumber\\
& &{}+\left(\frac{R_{\odot}}{R}v_{\phi}-v_{\odot}\right)\sin(l)\cos(b)
\label{vlostot}
\end{eqnarray}}
where $d$ is the heliocentric distance. Given $\rho(R,z)$ and ${\bf v}(R,z)$, the \hi\ column density $N_{\hi}(l, b, v)$ or the brightness temperature $T_{\rm B}(l,b,v)$ can be easily derived. We also included an isotropic velocity dispersion of $20 \kms$ for the \hi\ halo, spreading the computed components $N_{\hi}(l, b, v)$ over the neighboring channels.

\subsection{Toy models}\label{toylv}

\begin{figure*}[t]
\centering
\includegraphics[trim = 0mm 100mm 0mm 0mm, width =  \textwidth] {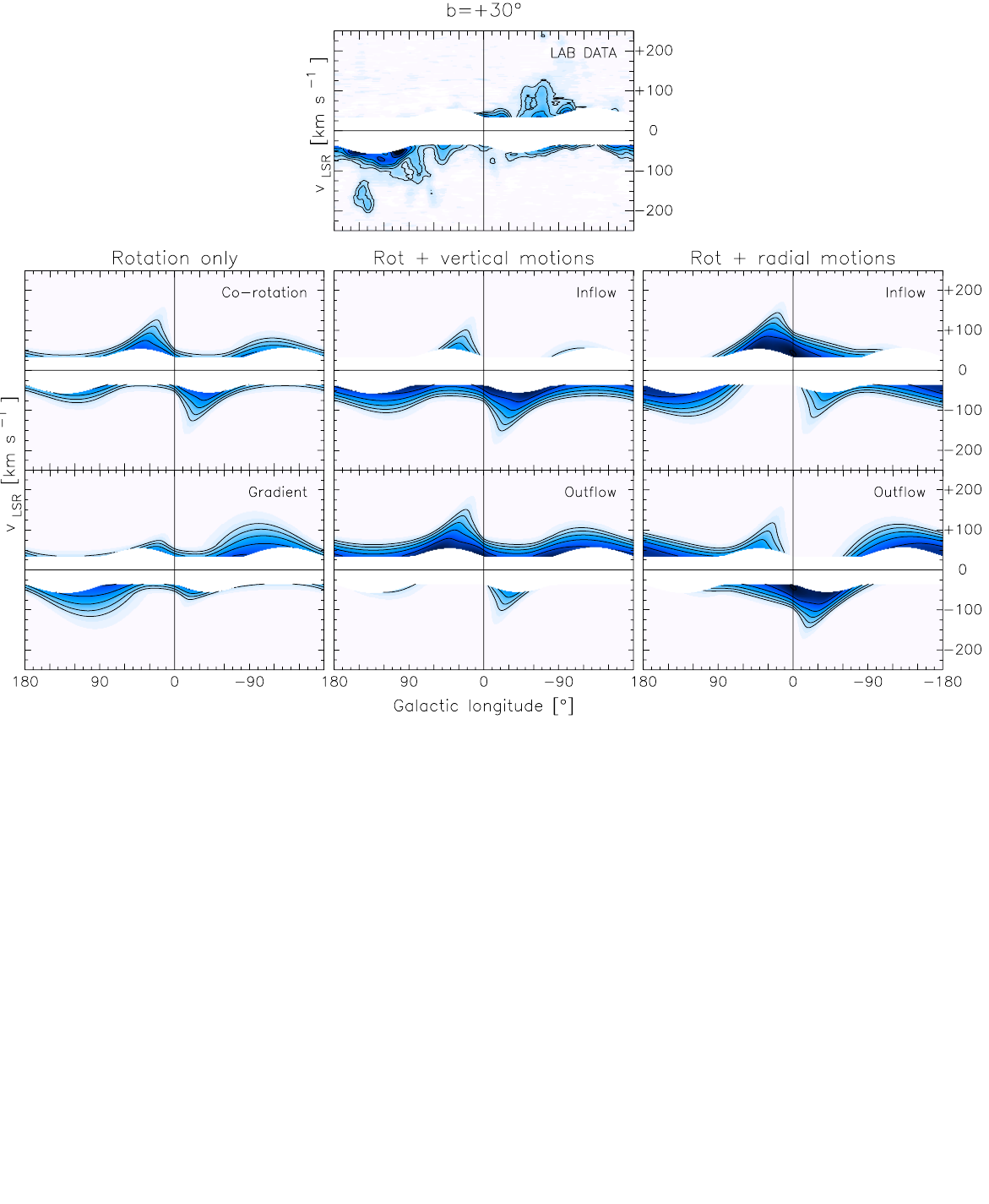}
\caption{$l\!-\!v$ slices at $b=30\de$ for the LAB dataset (top panel) and for six different halo models. \emph{Left panels}: halo corotating with the disk (top) and rotating with vertical lag $dv_{\phi}/d|z| = -20 \kmskpc$ (bottom); \emph{central panels}: halo corotating with vertical motion $v_z=-50\kms$ (top) and $v_z=+50\kms$ (bottom); \emph{right panels}: halo corotating with radial motion $v_R=-50\kms$ (top) and $v_R=+50\kms$ (bottom). Regions with $|v_{\rm DEV}|<35\kms$ have been excluded. All the cubes have been smoothed to $8\de$ resolution. Contour levels are: $40$, $80$, $160$, $320$, $640$, $1280\rm{\ mK}$ in brightness temperature.}\label{toy}
\end{figure*}
\begin{figure*}[bth!]
\includegraphics[trim = 0mm 185mm 0mm 0mm, width = \textwidth] {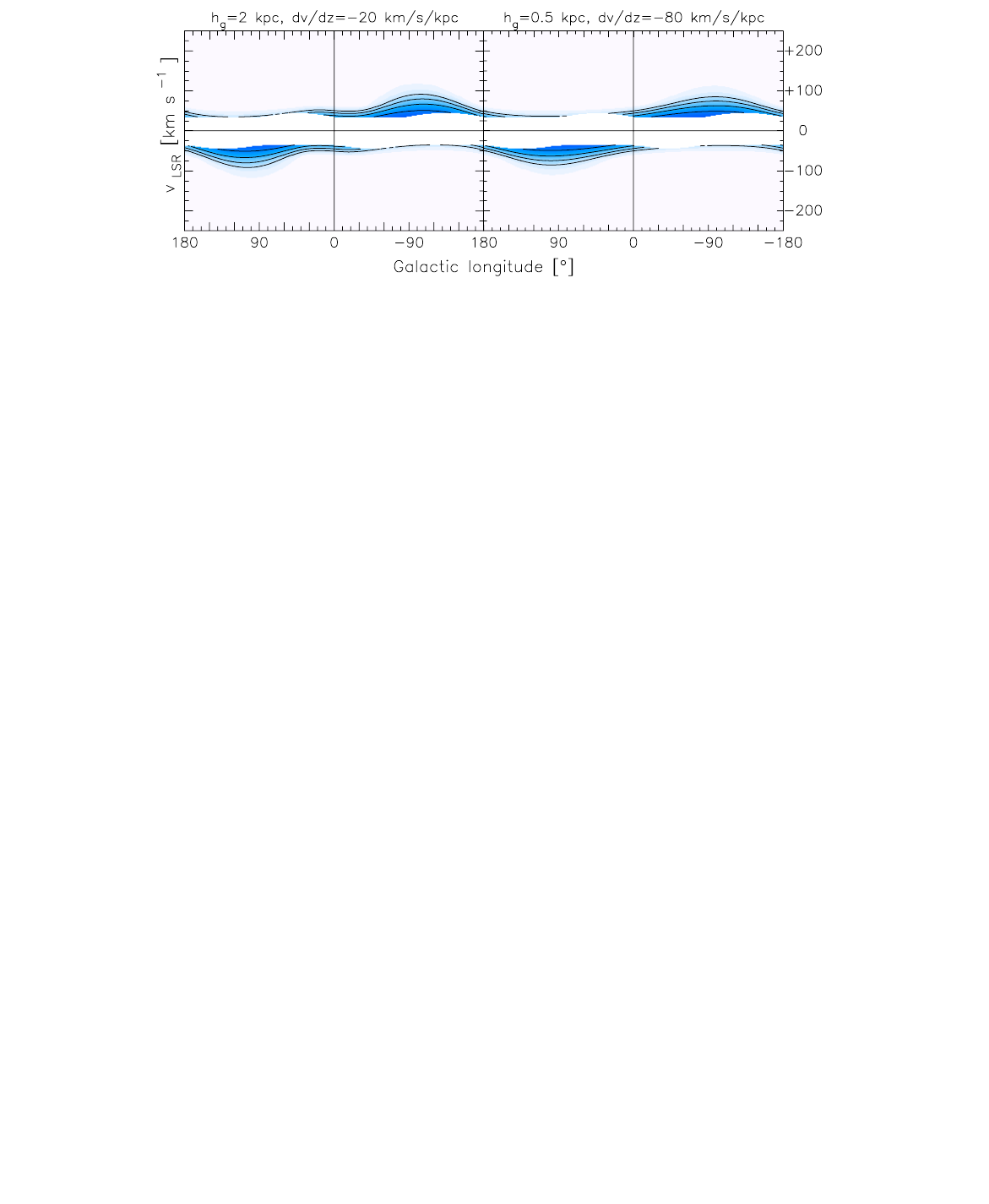}
\caption{The effect of the gradient-thickness degeneracy: the $l\!-\!v$ slices at $b=45\de$ for the two different rotating halo models are almost indistinguishable. The disk emission has been excluded and the cubes have been smoothed to $8\de$ resolution. Contour levels are: $40$, $80$, $160$, $320$, $640$, $1280\rm{\ mK}$ in brightness temperature.}\label{degen}
\end{figure*}

To better understand the effects of the different kinematic parameters on the pseudo-datacubes we first built some `toy' models with simple kinematics. For simplicity we set the \hi\ halo scale-height to $2\kpc$ and its mass to $5\times10^8\msun$. 
We removed the disk emission as discussed in \ref{disk}.
Figure \ref{toy} shows the longitude-velocity slice ($l\!-\!v$) at the galactic latitude $b = 30\de$ for the LAB data and for six different models. The data panel shows that after the disk cut in $v_{\rm DEV}$ (white region at low velocities) a large fraction of emission still remains.

A halo with the same kinematics as the disk (centre left panel) poorly reproduces the data: in this model the inner halo regions ($l\inÊ[0\de, 90\de]$ at $v_{\rm LOS}\!>\!0$ - the `receding' quadrant QI, and $l\inÊ[-90\de, 0\de]$ at $v_{\rm LOS}\!<\!0$ - the `approaching' quadrant QIV) show `horn-like' structures due to the fast rotation of the gas, whilst in the data these structures are absent. 
Also, the regions related to the external halo ($l\inÊ[0, 180\de]$ at $v_{\rm LOS}\!<\!0$ and $l\inÊ[-180, 0\de]$ at $v_{\rm LOS}\!>\!0$) do not reach the same \vlos of the data.
Adding a vertical lag in the halo rotation (bottom left panel) improves the comparison with the data: the horns disappear thanks to the lower rotation, while the external halo reaches higher velocities along the line of sight. 
This comparison suggests that the \hi\ halo of the Galaxy does not corotate with the disk.

A vertical motion (central panels) globally shifts the halo emission to negative $v_{\rm LOS}$ for $v_z\!<\!0$ or to positive $v_{\rm LOS}$ for $v_z\!>\!0$, because the gas, for any fixed $b$, is approaching the line of sight with an additional velocity $v_z \sin(b)$. 
This motion is not visible at low latitudes, but dominates at the highest ones. 
The high latitude emission in the LAB data occurs mostly at $v_{\rm LOS}\!<\!0$ suggesting the presence of a global vertical infall of the gas, as already pointed out by the analysis of the HVCs and IVCs kinematics \citep{Bajaja1987}. A global vertical outflow from the disk is excluded.

The effects of the radial motions (right panels) are not easily foreseen, because of the off-centred position of the Sun with respect to the rotation axis. Obviously around the Anti-Centre region ($l=180\de$) the emission is shifted toward negative (for $v_R\!<\!0$) or positive (for $v_R\!>\!0$) velocities, because the gas is almost directly approaching the Sun or receding from it. Around the Galactic Centre the $l\!-\!v$ profile is globally `deformed' (with respect to the corotating case) toward positive velocities for $v_R\!>\!0$ and toward negative ones for $v_R\!<\!0$.

This qualitative analysis suggests that the \hi\ halo of the Milky Way has two kinematical properties: lagging rotation and global infall. 
In principle we could quantify the various velocity components by comparing our models with the data. 
However these models have been built by assuming a value for the scale-height of the \hi\ halo and we found this to be a critical assumption as there is a \emph{degeneracy} between the vertical rotational gradient and the halo scale-height. 
Because of projection effects, halving $h_{\rm g}$ and doubling $dv_{\phi}/d|z|$ (or vice versa) distributes the emission over the channels in similar ways at the high latitudes.
As an example, Fig.\ \ref{degen} shows the $l\!-\!v$ slice at $b\!=\!45\de$ for two halo models with circular rotation: the first with $h_{\rm g}\!=\!2\kpc$ and $dv_{\phi}/d|z|\!=\!-20\kmskpc$, the second with $h_{\rm g}\!=\!0.5\kpc$ and $dv_{\phi}/d|z|\!=\!-80\kmskpc$. 
The total fluxes of the models have been normalized to the same value. 
It is evident that, even if the parameters used differ significantly, the first model is almost indistinguishable from the second.

Since at high latitudes we are not able to discriminate between a strongly lagging thin halo and a weakly lagging thick halo, we need an independent measurement for either $h_{\rm g}$ or $dv_{\phi}/d|z|$.
We tried to estimate $h_{\rm g}$ from the emission along $b$ in the \hi\ velocity-integrated maps (which do not depend on the kinematics) but we found that this method is also degenerate. Consequently, we searched for an independent measurement of the velocity gradient.

\section{The rotation velocity gradient in the inner Galaxy}\label{gradient}

\subsection{Method of tangent-point}\label{method}

The classical method to estimate the mid-plane rotation curve in the inner Galaxy (inside the solar circle) is that of the \emph{tangent-point}: if the gas is in circular rotation without turbulent motions, for any direction $l$ the highest velocity measured in the profile (the `terminal velocity' $v_{\rm ter}$) is the line-of-sight velocity of the ring at the tangent-point radius $R_{\rm tan}=R_{\odot}|\sin(l)|$ (the `tangent velocity' $v_{\rm tan}$). 
Therefore, using eq. (\ref{vlos}) for $b=0\de$ we have $v_{\phi}(R_{\rm tan}, 0) = |v_{\rm ter} + v_{\odot} \sin(l)|$. 
The presence of turbulent motions smears the \hi\ line profiles, so $v_{\rm ter}$ is actually lower than the highest velocity and it has to be searched in the wings of the profiles with some assumptions. 
Some authors used a brightness temperature threshold \citep{Malhotra1995}, others chose values related to first peak in the spectrum \citep{Kerr1962,Kerr1964} or to the integral of the profile past this peak \citep{Shane1966}. 
\citet{Levine2008} used a more complex approach based on the fitting of the whole line profile around the tangent velocity \citep{Celnik1979}. 
 
All those methods which make use of the end-line peak in the spectra - including that of \citet{Levine2008} - are not applicable at high latitudes since this peak, due to the falloff in the \hi\  density at high $z$, is too faint to be detected. 
Thus, we adopted a different approach, often used for the external galaxies: we assumed a certain velocity dispersion for the gas, and, for each line profile, we performed a Gaussian fit to the emission falloff leaving the amplitude and the mean ($v_{\rm ter}$) of the Gaussian as free parameters. This method, if applied at the mid-plane, gives similar results as most of the approaches described above, but it also has the advantage to be suitable at different latitudes (see below). In exchange it requires a higher signal-to-noise (S/N) ratio with respect to other approaches, since it uses for each spectrum only the last points at low brightness temperature. 
To improve the S/N we smoothed the datacubes: tests performed on the toy models (Section \ref{calibration}) showed that after smoothing to $3\de\!-\!4\de$ this approach works well for values of S/N typical of the LAB data.

To estimate the inner velocity gradient of the Milky Way we applied the tangent-point method outside the mid-plane. For any $(l,b)$ in the inner Galaxy, the tangent point is at $R_{\rm tan}=R_{\odot}|\sin(l)|$ and $z_{\rm tan} = R_{\odot}\cos(l)\tan(b)$. However, if the gas is not exactly in corotation at any height from the plane, $v_{\rm ter}$ is no longer equal to $v_{\rm tan}$. For example, in a lagging halo for any $b\!\neq\!0\de$ the highest velocity emission comes from a region closer to us than the tangent point, and if we use the method described we would assign to $v_{\phi}(R_{\rm tan}, z_{\rm tan})$ a value valid for a lower $z$: basically we would overestimate $v_{\phi}$. 
The question is whether this \emph{geometric bias} forbids the derivation of $v_{\phi}(R,z)$, or there is a window in the $(R,z)$-space where this effect is negligible and therefore:
\begin{equation}\label{vtp}
v_{\phi}(R_{\rm tan}, z_{\rm tan}) \simeq \left|\frac{v_{\rm ter}}{\cos(b)} + v_{\odot} \sin(l)\right|.
\end{equation} 
We built 3D models to check this (Section \ref{calibration}) and found that it is feasible to estimate the rotation up to a few kiloparsecs from the mid-plane. 

\subsection{Tests using the models}\label{calibration}

\begin{figure}[tb!]
\includegraphics[width=0.5\textwidth] {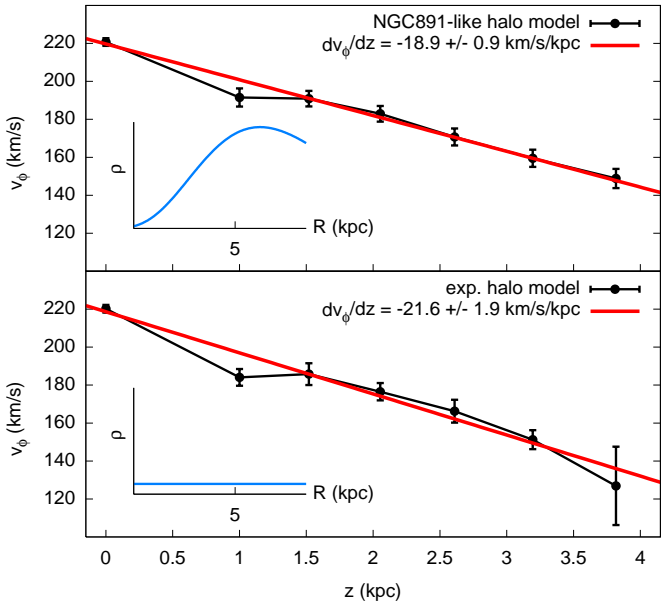}
\caption{Rotation velocity $v_{\phi}(z)$ (black points) averaged in the region $3.5\!<\!R\!<\!5.5\kpc$ computed with the tangent point method for the model A (top panel) and B (bottom panel). The red thick line is the linear fit. The insets show the radial density distribution from $R=0$ to $R_{\odot}$.}
\label{toygrad}
\end{figure}

We applied the tangent-point method to different models to verify that: a) in the presence of moderate gradients, the geometrical bias produces negligible errors and $dv_{\phi}/dz$ can be correctly derived; b) different density distributions weakly affect the estimate of the gradient. We built two pseudo-datacubes with the same \hi\ halo mass ($5\times10^8\msun$) and the same kinematics: $v_{\phi}(R,0) = 220\kms$ and $dv_{\phi}/d|z| = -20\kmskpc$. In the first (model A) the \hi\ density distribution follows eq. (\ref{halo891}), in the second (model B) the halo density decreases exponentially as $|z|$ increases and it is constant in $R$ (see insets in Fig.\ \ref{toygrad}). The vertical scale-height is $2\kpc$ for both models. We added a velocity dispersion of $20\kms$ for the halo gas.
  
We applied the method described in Section \ref{method} to both models, calculating $v_{\phi}(R,z)$ for $|z|\!<\!6\kpc$ using the input \hi\ velocity dispersion as the sigma of the Gaussian for fitting the profiles. In the absence of noise we are able to perform fits using components with an arbitrary low $T_{\rm B}$, taking into account only the dispersion falloff of the lines. In this ideal case the resulting velocity fields of both models are perfectly reconstructed for $1\!<\!R\!<\!5\kpc$ and $1\!<\!z\!<\!4\kpc$. Outside this window the geometric bias (see \ref{method}) is no longer negligible.

We added a Gaussian noise to our models, with a dispersion of $50\mK$ ($1\sigma_{\rm rms}$ of the Hanning-smoothed release of the LAB dataset), and we performed the fits using only the components with $T_{\rm b}\!>\!2.5\sigma_{\rm rms}\!=\!125\mK$: the velocity field resulted unsatisfactory for model B, showing that a higher S/N was required. To improve the S/N we smoothed the model cubes to $4\deg$ and performed a further Hanning Smoothing: this brought the average $1\sigma_{\rm rms}$ to $5\mK$ and the threshold of the fit to $12.5\mK$. We re-performed the fits averaging all the profiles within the new beam and using only independent points, finding that for any fixed $R$ between $2$ and $5\kpc$ the $v_{\phi}(z)$ curve could be fitted with a straight line in good agreement with the input slope. We averaged $v_{\phi}(z)$ over a range of $R$ in order to improve the statistics on the single measurements, at $4\de$ of resolution there are about $3$ independent points for $3.5\!<\!R\!<\!5.5\kpc$.

Figure \ref{toygrad} shows the $R$-averaged $v_{\phi}(z)$ curve derived for the two models with the method described. We also included a point for the mid-plane. We assumed that the error on the single measurement $v_{\phi}(z)$ is given by the squared sum of three terms: a) the rms of the $v_{\phi}(z)$ distribution over $3.5\!<\!R\!<\!5.5\kpc$; b) the maximum statistical error on the Gaussian fits computed in this distribution;  c) a `base' error of $4\kms$ that takes into account uncertainties in the gas velocity dispersion. 
Fitting the points with a straight line we obtained $dv_{\phi}/d|z| = -18.9\pm0.9\kmskpc$ for model A and $-21.6\pm1.9\kmskpc$ for model B, in good agreement with the input gradient. 
We tested further this method varying the vertical scale-height of model A from $0.75\kpc$ to $3\kpc$, obtaining similar results. Our method depends weakly on the density distribution, so we can proceed to apply it to the LAB data.

\subsection{Application to the data} \label{application}

The \hi\ distribution in the Milky Way is neither exactly axisymmetric nor $b$-symmetric. Therefore we derived four $v_{\phi}(z)$ curves in the inner Galaxy ($R\!<\!R_{\odot}$), one for each quadrant (QI and QIV, see Section \ref{toylv}) and for the positive and the negative latitudes separately. We used the LAB dataset smoothed both spatially ($4\de$) and in velocity (Hanning), averaging the profiles within the new beam and using only the independent points. 
The threshold used is $2.5\times6\mK$ for QI and $2.5\times4\mK$ for QIV, since the noises in regions at positive and negative declinations slightly differ.
We assumed a velocity dispersion of $20\kms$ for the halo gas. This value has been estimated for NGC\,891 by \citet{Oosterloo2007}. In the Milky Way, \citet{Ford2008} found a cloud-to-cloud velocity dispersion of $18\kms$ for the \hi\  emission in the lower halo regions. 
The errors on the single fits were derived as described in Section \ref{calibration}.

We found that in the LAB Survey the statistical mean of the noise is not zero but instead systematically around $+10\mK$. In the full resolution dataset the $2.5\sigma_{\rm rms}$ threshold is well above these values, but after a $4\de$ smoothing has been applied it becomes of the same order. The nature of this emission is not clear, and systematic errors at this low level in the LAB Survey may exist (P. Kalberla, private communication). We preferred to re-set the noise zero point channel by channel, subtracting the average value estimated in regions not contaminated by the Galactic emission. This is equivalent to using for our fit a threshold higher than $2.5\sigma$ different for each channel map. 

\begin{figure}[t]
\includegraphics[width=0.5\textwidth] {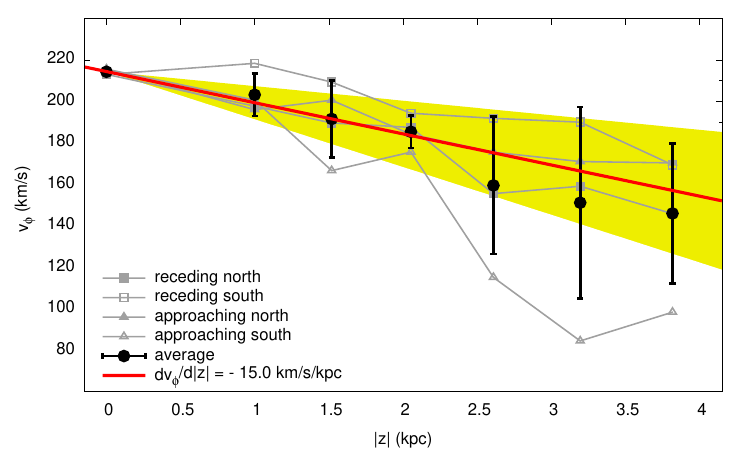}
\caption{Vertical rotational gradient in the inner Milky Way. \emph{Thin lines}: rotation velocity $v_{\phi}(z)$ averaged in the region $3.5\!<\!R\!<\!5.5\kpc$ derived with the LAB dataset using the tangent point method. Each curve stands for a different region. Error bars are not shown. \emph{Black points}: average $v_{\phi}(z)$ curve. \emph{Thick red line}: linear fit to the average curve. The yellow region stands for the $2\sigma$ error on the gradient ($1\sigma = 4\kmskpc$).}
\label{labgrad}
\end{figure}

Figure \ref{labgrad} shows the four $R$-averaged $v_{\phi}(z)$ curves derived for each quadrant, as well as the mean curve and its linear fit. We added two points belonging to the mid-plane, one for QI and one for QIV, averaged in $3.5\!<\!R\!<\!5.5\kpc$. We estimated the rotation curve in the mid-plane $v_{\phi}(R,0)$ (see thick lines in Fig.\ \ref{rc}) using the LAB data smoothed to $2\de$, a threshold of $2\K$ and $2\kms$ for the `base' error. The linear fit gives $-15\kmskpc$ for the average vertical falloff in the rotation velocity. We estimated an error of $4\kmskpc$ obtained using as a constrain the point around $z=2\kpc$; the shaded area in Fig.\ \ref{labgrad} shows the $2\sigma$ confidence level.

If vertical and radial motions are present they may affect the calculation of $v_{\phi}(z)$, since we showed (Section \ref{toylv}) that non-circular motions modify the shape of the $l\!-\!v$ distribution. We checked this by adding a vertical infall ($v_z = -40\kms$) and a radial inflow ($v_R = -30\kms$) to our models and re-estimating $v_{\phi}(z)$. We found that in the presence of these motions the gradient would be overestimated in the QI regions and underestimated in the QIV regions. 
However, the overall effect is inside the error bar of our gradient estimate, and we can neglect it.

Recently it has been pointed out that the velocity of the solar circle $v_{\odot}$ could be significantly higher than the standard value of $220\kms$ \citep{Reid2009}. We note that the value of $v_{\odot}$ has no impact on our gradient estimate. Eq.\ \ref{vtp} states that changing $v_{\odot}$ would shift up or down the rotation velocity at a certain longitude (i.e. at a certain $R$) for any $z$. For instance, in Fig.\ \ref{labgrad} all the points would be vertically shifted by the same amount, with no consequences for the fitted slope.

Finally, we point out that the implicit assumption behind the method used is that in the regions investigated the gas is dense enough to show a significant emission. Testing the method with very different density distributions and scale-heights (Section \ref{calibration}) gave comforting results. However, if an unusually deep depletion of \hi\ were present in the inner Galactic regions our gradient would be overestimated.
 
\section{Results}\label{results}

In the previous Section we derived the inner rotational gradient of the Milky Way \hi\ halo. We proceed now to estimate the remaining three free parameters of the halo: the vertical scale height ($h_{\rm g}$), the vertical component ($v_z$) of the global motion and the radial one ($v_R$).
Given the inapplicability of the tangent-point method we can not derive the rotation velocity for $R\!>\!R_{\odot}$. Thus we assume that: a) the mid-plane rotation curve remains flat beyond $R_{\odot}$ (see fig.\ \ref{rc}); b) the rotational gradient, derived for $3.5\!<\!R\!<\!5.5\kpc$, is the same ($-15\kmskpc$) at each radius.

\subsection{Minimization of residuals}\label{residuals}
We built a series of models with different combinations of the three parameters ($h_{\rm g}$, $v_z$, $v_R$) and searched for the values that minimize the differences with the data. We explored the following ranges: from $-100\kms$ to $0\kms$ in steps of $10\kms$ for $v_z$, from $-50\kms$ to $50\kms$ in steps of $10\kms$ for $v_R$, from $0.5\kpc$ to $4\kpc$ in steps of $0.25\kpc$ for $h_{\rm g}(R)$. 
We excluded a positive range for $v_z$ since the data clearly do not show a systematic shift of the emission towards the positive velocities. 
For each model the total flux computed in the halo region was normalized to the observed halo flux by multiplying each $T_{\rm B}(l,b,v_{\rm LOS})$ for an appropriate factor, and since $T_{\rm B}\propto N_{\hi} \propto M_{\hi}^{\rm halo}$ this is a correction to the \hi\ halo mass. 
The resulting halo masses above $2$ disk scale-heights for different models vary from $0.5\times10^8 M_{\odot}$ to $9\times10^8 M_{\odot}$.

For each model we computed the difference with the LAB datacube pixel by pixel, and we summed these values using a $\cos(b)$ factor to take into account the projection effects. We used both a quadratic sum and an absolute sum, obtaining little differences. The sum were extended to all the components with the exception of the regions occupied by the disk emission ($|v_{\rm DEV}|\!<\!35\kms$) and by: a) the Magellanic Stream; b) the Leading Arm; c) the GCP complex \citep[or `Smith Cloud';][]{Smith1963} which is an isolated cloud with an anomalous kinematics \citep{Lockman2008}; d) the Outer Arm (see Section \ref{disk}); e) the external galaxies. We also disregarded components with $|v_{\rm LOS}|\!>\!250\kms$, since at these velocities only some isolated `Very High Velocity Clouds' contribute to the emission, and obviously our models do not attempt to reproduce these features. Furthermore, we excluded the QI and QIV regions for $|b|\!<\!50\de$ (the `horn-like' structure, see Section \ref{toylv}) since their emission - for a fixed density distribution - gives information mainly on $dv_{\phi}/d|z|$, which has already been set.

 \subsection{The best model}\label{bestmodel}

\begin{figure*}[htbp!]
\centering
\includegraphics[scale = 0.95]{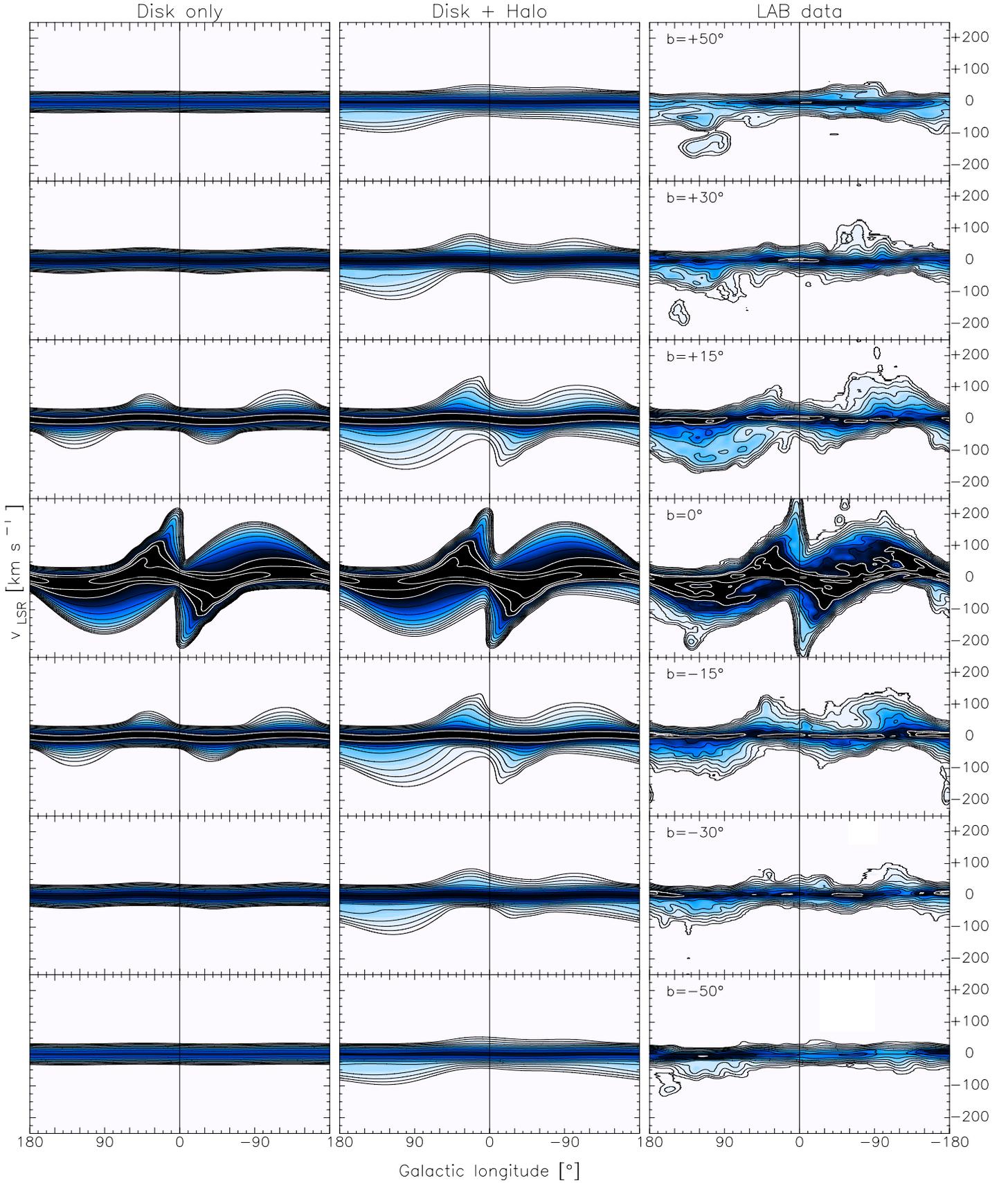}
\caption{$l\!-\!v$ slices at 7 different latitudes indicated at the top left corners of the rightmost plots. 
\emph{First column}: the \hi\ disk model; 
\emph{second column}: \hi\ disk + best halo model; 
\emph{third column}: the LAB data. 
Each datacube is smoothed to $8\de$ resolution. 
Contour levels in brightness temperature range from $0.04\K$ to $81.92\K$ scaling by a factor $2$.}
\label{lvbest0}
\end{figure*}

We found that the kinematical parameters that minimize the differences between the models and the data are $v_z\!=\! -20\kms$, $v_R\!=\! -30\kms$. For the vertical scale-height we obtained $1.75\kpc$ computing the residuals using the squared sum and $1.5\kpc$ using the absolute sum, thus we set $h_{\rm g}\!=\!1.6\kpc$. The \hi\ halo mass above $2$ disk scale-heights (see Section \ref{disk}) is $3.2\times10^8 M_{\odot}$. Table \ref{bmparameters} summarizes the parameters of the best halo model, as well as their errors. These latter were obtained by searching for `acceptable models' around the best set of parameters. Note that the \hi\ halo mass reported corresponds to $5\!-\!10\%$ of the total Galactic \hi.

\begin{figure*}[htbp!]
\includegraphics[trim = 0mm 158mm 0mm 0mm, width = \textwidth] {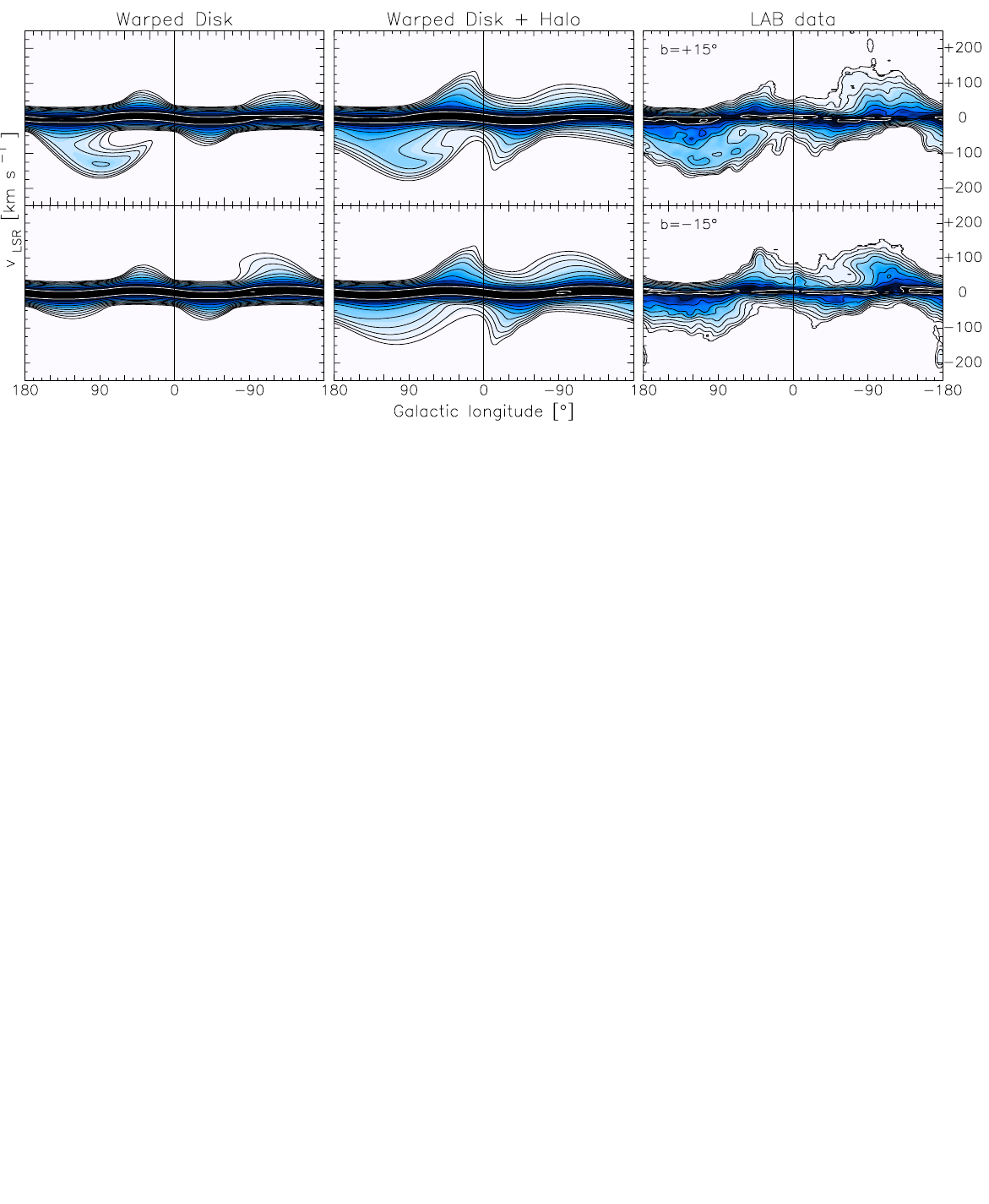}
\caption{$l\!-\!v$ slices at $b\!=\!+15\de$ (top) and $b\!=\!-15\de$ (bottom). 
\emph{First column}: the warped \hi\ disk model; 
\emph{second column}: the warped best model (\hi\ disk + halo); 
\emph{third column}: the LAB data. 
Each datacube is smoothed to $8\de$ resolution. 
Contour levels in brightness temperature range from $0.04\K$ to $81.92\K$ scaling by a factor $2$.}
\label{warpmod}
\end{figure*}

Figure \ref{lvbest0} shows seven representative $l\!-\!v$ slices at various latitudes for three datacubes: a model with only the \hi\ disk component, a model with both (disk + best halo) components, and the data. 
The \hi\ distribution in the disk was modelled as reported in KD08, but we used the rotation curve found in Section \ref{application} (thick lines in Fig.\ \ref{rc}), we added an isotropic velocity dispersion of $10\kms$ and we included an exponential decrease in density for $R\!<\!4.75\kpc$. 
We did not include the warp (but see Section \ref{refine}). 
The comparison between models and data clearly highlights how the \hi\ disk emission alone fails to reproduce the observations, especially at the intermediate latitudes ($b\!=\!\pm15\de$ and $b\!=\!\pm30\de$). 
Instead, adding the halo component allows us to reproduce most of the \hi\ features at low brightness temperature ($T_{\rm B}\!\lesssim\!0.5\K$), especially those coming from the external regions ($0\de\!<\!l\!<\!180\de$ at $v_{\rm LOS}\!<\!0$, and $-180\de\!<\!l\!<\!0\de$ at $v_{\rm LOS}\!>\!0$) at latitudes $|b|\!\ge\!15\de$. 
The inner regions (QI and QIV) appear to reach velocities higher than the data: this can be interpreted as evidence of a larger gradient in the innermost regions - as found by \citet{Oosterloo2007} for NGC\,891 - or of a depletion of \hi\ in these areas (see next Section).
Also, the LAB data shows asymmetries between the positive and the negative latitudes that we can not reproduce with our symmetric models.

In order to estimate the impact of the local \hi\ emission on the above parameters we performed the following test.
We considered three models, each one excluding a different part of the halo: in model \#1 we included only the inner halo by excluding the \hi\ emission at $R\!>\!8\kpc$, in model \#2 we added the local gas by excluding the emission at $R\!>\!9\kpc$, in model \#3 we excluded the \hi\  at $R<9\kpc$ in order to take into account only the outer emission. For each model we performed the analysis described in Section \ref{residuals}. We found that the best parameters for model \#2 are the same as those of table \ref{bmparameters}, while those of model \#1 and \#3 slightly differ: we obtained ($v_z=-25\kms, v_R=-30\kms, h_{\rm g}=1.1\kpc$) for the first and ($v_z=-15\kms, v_R=-20\kms, h_{\rm g}=1.6\kpc$) for the latter. These values are all consistent within the errors with those of table \ref{bmparameters} and we can conclude that our results are global.

To understand if a variation of the parameters $R_{\rm g}$ and $\gamma$ affects our results, we build four new models using respectively $R_{\rm g}=2.4\kpc$, $R_{\rm g}=1.1\kpc$, $\gamma=2$ and $\gamma=8$ (see eq.\,\ref{halo891}). For each model we re-performed the residuals minimization, finding that in all cases the best parameters are consistent within the errors with those of table \ref{bmparameters}.
Finally, we investigated the impact of $v_{\odot}$ on the residuals by re-building all models using $v_{\odot}=250\kms$ for both the mid-plane rotation and the velocity projections, obtaining no variation from the above best parameters.

\begin{table}[t]
\centering
\begin{tabular}{lrl}
\hline
\hline
& & \\
\hi\ mass & $3.2^{+1.0}_{-0.9}\times10^8$ & $ M_{\odot}$ \\
& &\\
Scale-height ($h_{\rm g}$) & $1.6_{-0.4}^{+0.6}$&$ \kpc$\\
& & \\
Rotational gradient ($dv_{\phi}/d|z|$) & $-15^{+4}_{-4}$&$ \kmskpc$\\
& &\\
Vertical velocity ($v_z$) & $-20^{+5}_{-7}$&$ \kms$ \\
& &\\
Radial velocity ($v_R$) & $-30^{+7}_{-5}$&$ \kms$ \\
& &\\
\hline
\end{tabular}
\caption{Global parameters of the \hi\ halo of the Milky Way, from our best halo model (see text).}
\label{bmparameters}
\end{table} 

\subsection{Further refinements}\label{refine}

We considered the effect of a Galactic warp. Following \citet{Levine2006}, we modeled the offset of the mid-plane $z_0$ from the plane $z\!=\!0$ with the formula
\begin{equation}\label{warp}
z_0(R,\phi) = W_0(R) + W_1(R)\sin(\phi-\phi_1(R)) + W_2(R)\sin(2\phi-\phi_2(R))
\end{equation}
where the dependencies on $R$ of the three amplitudes $W_i$ and the two phases $\phi_j$ have been taken from \citet[][Fig.\ 16 and Fig.\ 17 therein]{Kalberla2007}. We considered the warp only for $R\!>\!R_{\odot}$, and we set $W_2(R)\!=\!0\kpc$ for $R\!<\!16\kpc$. 
Under the assumption that the \hi\ halo stratifies like the \hi\ disk, we included the warp in our models just substituting the density computed in $(R,z)$ with that computed in $(R, z-z_0)$ for both the disk and the halo distribution. 
Also, the halo rotation curve $v_{\phi}(R,z)$ has been substituted with $v_{\phi}(R, z-z_0)$. 
This can not be considered a full treatment of the warp effect since we used eq. (\ref{vlostot}) to compute the line of sight velocity and we assumed that the orbits are circular. However, the mid-plane significantly deviates from the $z=0$ plane only at large radii ($R\!>\!15\kpc$) and around $\phi\sim80\de$, which makes the contribution of the halo emission in the warp region negligible.

Figure \ref{warpmod} shows the two $l-v$ plots at $b= \pm 15\de$ for the data (rightmost panels) compared with a disk-only model (left column) and a disk$+$halo model, both with the inclusion of the warp.
The warp is very asymmetric and mostly visible at positive latitudes for $l=90\de$.
In this region - the Outer Arm zone - the \hi\ emission is deformed towards negative velocities and the model reproduces the data significantly better. 
At higher latitudes the effect of the warp is negligible and the other $l-v$ diagrams in Fig.\ \ref{lvbest0} remain unchanged.
Clearly, the inclusion of a warp does not significantly affect our results for the \hi\ halo.

\begin{figure}[tbp!]
\centering
\includegraphics[trim = 30mm 20mm 10mm 20mm, width=0.5\textwidth]{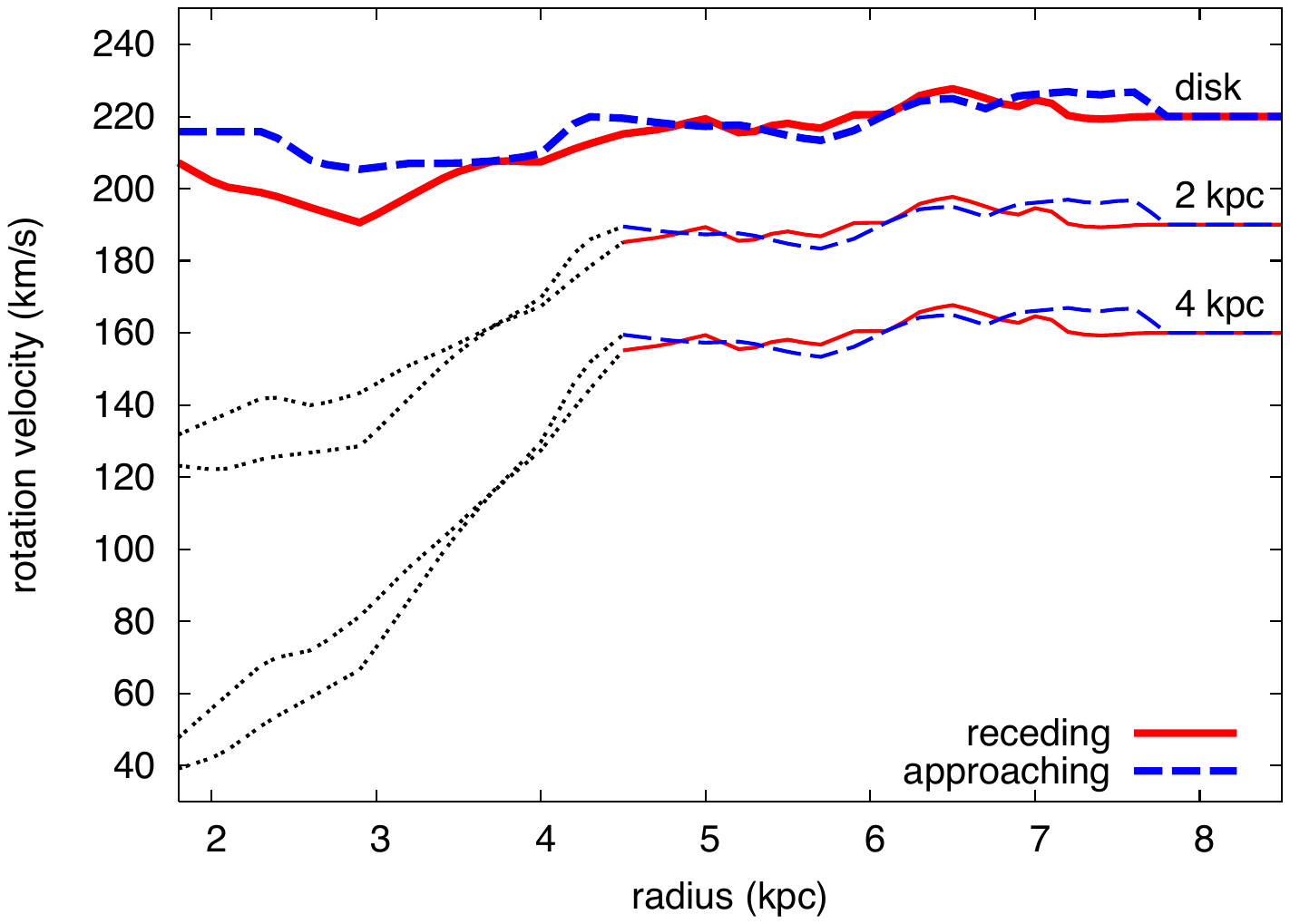}
\caption{Rotation curves as a function of the height from the Milky Way mid-plane used in modelling. The thick lines on top stand for the mid-plane rotation. \emph{Red solid lines}: receding curves; \emph{blue dashed lines}: approaching curves; \emph{black dotted lines}: rotation velocities assumed for the `shallow rise' model at $R\!<\!4.5\kpc$. For $R\!>\!7.7\kpc$ we assumed that the curves are flat. }
\label{rc}
\end{figure}

\begin{figure*}[htbp!]
\includegraphics[trim = 0mm 160mm 0mm 3mm, width = \textwidth] {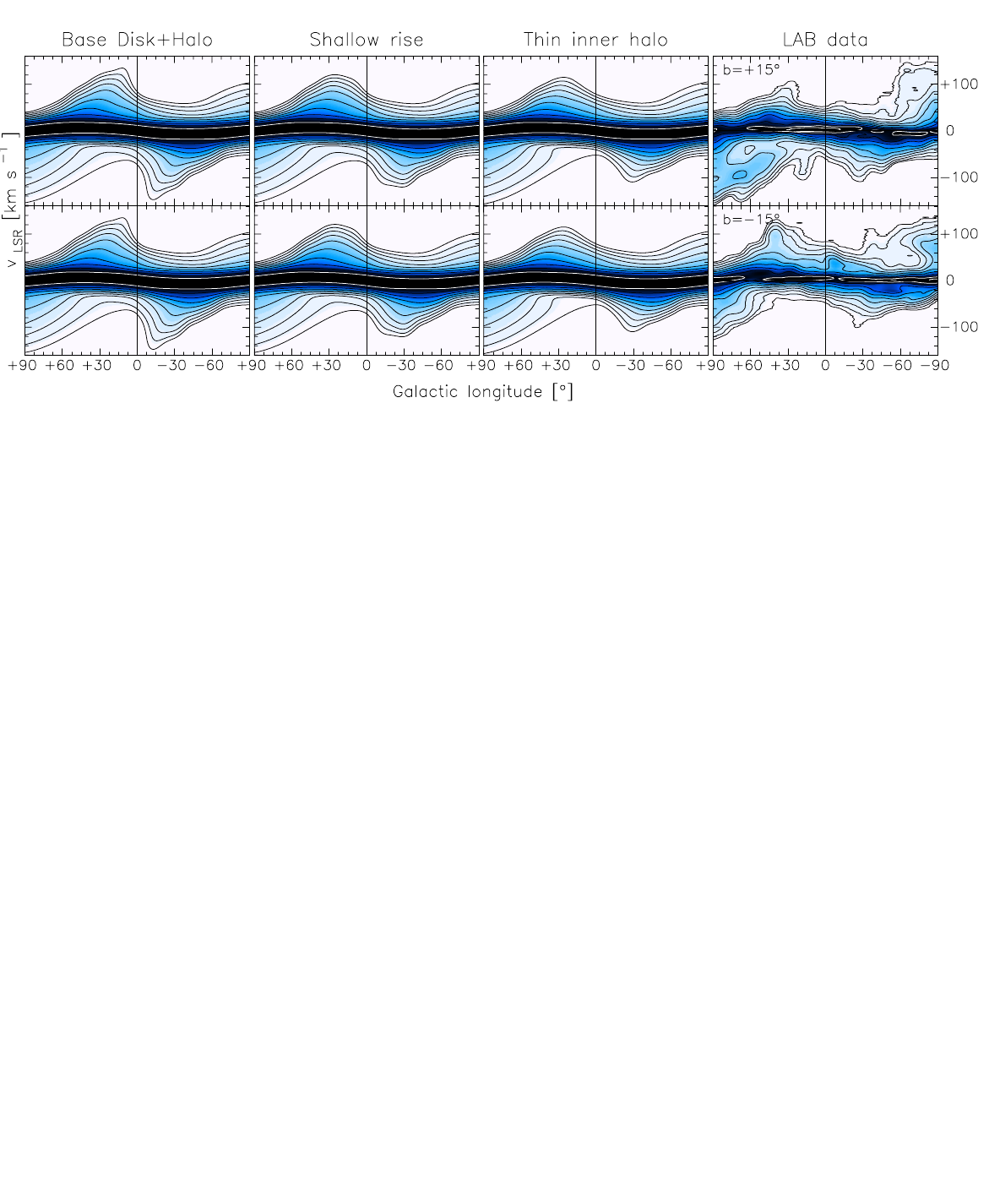}
\caption{$l\!-\!v$ slices at $b\!=\!+15\de$ (top) and $b\!=\!-15\de$ (bottom) at longitudes $-90\de\!<\!l\!<\!90\de$. 
\emph{First column}: the best model shown in Fig.\ \ref{lvbest0}; 
\emph{second column}: the model refined including a radial dependence of $dv_{\phi}/d|z|$ for $R\!<\!4.5\kpc$; 
\emph{third column}: the model refined including a linear decrease of the halo thickness for $R\!<\!R_{\odot}$; 
\emph{fourth column}: the LAB data. 
Each datacube is smoothed to $8\de$ resolution. Contour levels in brightness temperature from $0.04\K$ to $81.92\K$ scaling with a factor $2$.}\label{shallowmod}
\end{figure*}

In order to reproduce the emission related to the inner regions (QI and QIV) we considered two effects separately. 
The first is the inclusion of a radial dependence for the rotational gradient in the inner halo.
We assumed that for $R\!<\!4.5\kpc$ the magnitude of the rotational lag increases linearly each kiloparsec by $10\kmskpc$. 
The effect of this `shallow rising' rotation curve (dotted curves in Fig.\ \ref{rc}) is to lower $v_{\rm LOS}$ in the inner halo, reducing the `horns' (see Section \ref{toylv}) in the $l\!-\!v$ diagrams. 
Figure \ref{shallowmod} shows that this effect is not sufficient, since the $l\!-\!v$ profiles obtained for this model (second column) differ only slightly from the one with a constant vertical velocity gradient (first column).

As a second attempt we reduced the halo thickness in inner regions. 
We assumed that the vertical scale-height of the \hi\ halo decreases linearly for $R\!<\!R_{\odot}$, vanishing at the Galactic Centre. 
Figure \ref{shallowmod} shows a model with this thin inner halo (third column) and constant rotational gradient of $-15 \kmskpc$. 
This model reproduces fairly well the \hi\ emission in these regions of the datacube.

We conclude that the scale-height of the \hi\ halo of the Milky Way inside $R_{\odot}$ is likely to be lower that our average value of $1.6 \kpc$, the rotational gradient may also be larger than $15 \kmskpc$ in the inner parts. It is difficult to disentangle these two effects in our analysis. For instance, given the degeneracy between $dv_{\phi}/dz$ and $h_{\rm g}$ (Section \ref{toylv}), a non-constant gradient with R would imply a corresponding variation of the vertical scale-height.
Both a decreasing scale-height and an increasing gradient in the inner parts are consistent with a galactic fountain origin for the Milky Way halo (see Section \ref{interpretation}).

\section{Discussion}\label{discussion}

In this paper we have estimated the global properties of the Milky Way \hi\ halo by building 3D models and comparing them with the LAB Survey dataset. 
We found that the Galactic \hi\ halo can be globally described by the following average kinematical parameters: $dv_{\phi}/d|z|\!=\!-15\pm4\kmskpc$, $v_z\!=\!-20^{+5}_{-7}\kms$ and $v_R\!=\!-30^{+7}_{-5}\kms$ (Section \ref{bestmodel}). 
In the following we discuss the results obtained.

\subsection{Physical interpretation}\label{interpretation}

\citet{FB06} successfully reproduced the vertical gaseous distribution of NGC\,891 and NGC\,2403 by modeling the halo as a continuous flow of non-interacting galactic fountain clouds moving ballistically in the galactic potential \citep[see also][]{Collins2002,Heald2007}.
However they were forced to include interactions of these clouds with the environment in order to reproduce the vertical rotational lag \citep{FB08}: in this scenario the fountain clouds accrete gas from an ambient medium with a low angular momentum with respect to the galactic disk. 
This process slows down the rotation of the galactic fountain clouds and produces gas accretion towards the disk. 
Hydrodynamical simulations show that the Galactic corona could be the reservoir of this accreting gas \citep{Marinacci2010a}. 
The typical orbits of this `accreting galactic fountain'  differ from those of a fountain without accretion for the presence of inward radial motions \citep[Fig.\ 6 in][]{FB08} which indeed have been observed in the \hi\ halo of NGC\,2403 \citep{Fraternali2001}. 
In Section \ref{bestmodel} we showed that the \hi\ halo of the Milky Way has global vertical and radial motions of inflow, and in Section \ref{application} we estimated a vertical lag equal to the one found in NGC\,891. 
These results suggest that the dynamical interpretation given for the gaseous halo of NGC\,891 is suitable for the Milky Way halo as well.

The vertical inflow motion estimated ($v_z\!<\!0$) can be explained supposing that the galactic fountain gas is initially ionized for a certain part of the rising orbit, becoming visible in \hi\ only after recombination. 
If one considers a typical fountain orbit where the gas is ejected vertically from the disk with an ejection velocity $v_{\rm kick}$ and a vertical velocity that decreases roughly linearly with time \citep[Fig.\ 6 in][]{FB08}, to have an average vertical velocity $\overline{v_z}\!<\!0$ the gas should be ionized for a fraction $-2\,\overline{v_z}/v_{\rm kick}$ of the rising orbit. 
This fraction is around $50\!-\!70\%$ for the vertical velocity derived in Section \ref{bestmodel} ($-20^{+5}_{-7}\kms$) and for typical ejection velocities of $60\!-\!80\kms$. 

The negative value of $v_R$ in the halo is more difficult to interpret. Note that, in the context of a phase-change fountain, $v_R\!<\!0$ does not imply a net inflow for the whole gaseous halo but only for the neutral descending clouds. For each cloud, this inward motion would be roughly balanced by an outward motion during the first part of the orbit, when the gas is ionised. A ballistic non-interacting fountain cloud may have some negative $v_R$, but only at the very end of its orbit \citep{FB06}. It seems therefore more likely that such a large value of $v_R$ is produced by the loss of angular momentum due to gas accretion hypothesized by \citet{FB08}. 
In order to fully understand these effect, one requires a full dynamical modelling of the halo, which will be the subject of further investigations. 

\subsection{Relation to previous works}

\citet{Lockman1984} found that in the lower halo regions of the inner Galaxy a large amount of \hi\ is present. 
He estimated that this medium had a vertical scale-height of $500\pc$ approximately constant for $4\!<\!R\!<\!8\kpc$. Also, \citet{Lockman2002} confirmed that the diffuse \hi\  emission in these areas is organized into discrete clouds \citep[see also][]{Ford2008}.
In Section \ref{bestmodel} we found that all the acceptable models have larger scale-heights.
The difference can be explained as follows: 
a) our average scale-height is derived using the emission from the entire halo while it seems in the innermost regions of the Galaxy the halo is thinner (Section \ref{bestmodel} and \ref{refine});
b) Lockman derived the position ($R, z$) of the emission at the tangent point assuming co-rotation for the halo gas, and he pointed out that in case of a lagging halo the amount of \hi\ would be underestimated. 

\citet{Levine2008} were the first to derive a value for the vertical falloff in the Galactic rotation velocity using a direct approach. They applied the tangent point method to both the VLA Galactic Plane Survey \citep{Stil2006} and Southern Galactic Plane Survey \citep{McClureGriffiths2005} dataset to derive the inner rotation curves up to $|z|\!=\!100\pc$. 
The method used to determine the tangent velocity was based on fitting the whole profile around $v_{\rm ter}$ using a form derived by \citet{Celnik1979}. 
They found an average lag of $-22\pm6\kmskpc$, in agreement with our estimate obtained in the halo region for $1\!<\!|z|\!<\!4\kpc$. 
With the resolution and the S/N of the LAB data we can not determine the rotation curve for $|z|\lesssim1\kpc$, thus a direct comparison with the work of Levine et al.\ is not possible.

A local value for the vertical rotational gradient was derived by \citet{Pidopryhora2007} in a small region (the `plume') above the Ophiucus superbubble. 
Assuming that the feature is located at the tangent point, they found a lag of $-27\kms$ with respect to the corotation at position $R=4\kpc$, $z=3.4\kpc$. This means an average gradient of $-8\kmskpc$ at that radius. 
In the same region we measured a gradient of roughly a factor $2$ larger (see receding northern curve in Fig.\ \ref{labgrad}).
In the LAB data smoothed at $4\de$, the Ophiucus superbubble is not visible and we are likely measuring the average velocity of the surrounding region where the superbubble is expanding.


In the attempt to reconstruct the spatial \hi\ distribution by a 3D de-projection of the LAB data, rotation curves at several heights from the mid-plane have been derived by \citet{Kalberla2007} using a theoretical approach, previously adopted by \citet{Levine2006}. 
They modelled the Galactic dark matter distribution so that the resulting rotation curve (obtained assuming that the gravity dominates the gas dynamics everywhere) gives an hydrostatic \hi\ distribution whose properties (e.g. the flaring) are consistent with the dark halo itself. 
The curves obtained \citep[Fig.\ 9 in][]{Kalberla2007} differ from those we derived (Fig.\ \ref{rc}), specifically the vertical lag at $R\!=\!4.5$ averaged for $1\!<\!|z|\!<\!4\kpc$ is around $-22\kmskpc$ and it decreases as $R$ increases. 
Our approach does not rely on any assumption about the dynamical state of the system as we simply derive the best model that reproduces the observations.
We notice that if the \hi\ halo is produced by galactic fountains interacting with the environment (see Section \ref{interpretation}), assuming that the dynamics of the gas is dominated by gravity is not necessarily a good description of the system \citep{FB08}. 
\citet{Marinacci2010b} also showed that for NGC\,891 a hydrostatic model does not reproduce satisfactorily the data.

Finally, KD08 analysed the Galactic \hi\ model derived in \citet{Kalberla2007}. 
They found that $\sim10\%$ of the \hi\ is not related to the disk, in agreement with our estimate for the \hi\ halo mass (Section \ref{bestmodel}).

\subsection{HVCs and IVCs}\label{HVCsIVCs}

HVCs and IVCs are regarded as the main evidence for the presence of a gaseous halo in our Galaxy, and it is important to understand how they fit with the results of this work. 
Even if the emission of these objects dominates the \hi\ flux in regions outside the \hi\ disk, at low column densities the sky is basically filled with neutral hydrogen emission (Fig.\ \ref{allskyhalo}). 

Distance estimates for the IVCs show that they are a local phenomenon \citep[][and references therein]{Wakker2004} and their disk-like metallicities and kinematics suggest a galactic fountain origin. 
In this paper we showed that these clouds are embedded in a diffuse halo, which shares the same kinematics.
The emission of this gaseous component (IVCs + diffuse \hi) is reproduced by our model.
This suggests that the whole \hi\ halo is made up of IVC-like objects, most of which are at large distances and thus unresolved.
If these clouds had an average \hi\ mass of $\sim10^5M_{\odot}$, the \hi\ halo would contain $3\!-\!4$ thousand such objects.

Recent studies on HVCs established that these objects have higher distances and heights from the plane and lower metallicities than IVCs \citep{Wakker2007,Wakker2008}.
In general our halo model does not reproduce HVCs and the halo parameters derived in Section \ref{bestmodel} are not representative of their properties.
This suggests that the nature and origin of these clouds are different from that of the halo described here.
There are special cases such as the the well-known complexes C and A, visible in Fig.\ \ref{lvbest0} at $b=+50\de$ and $v_{\rm LOS}\!\sim\!-140\kms$ and $b=+30\de$ and $v_{\rm LOS}\!\sim\!-160\kms$.
Although separated from the `normal' halo gas these clouds appear in a location that is not completely peculiar.
They follow the trend of the halo emission but at a more extreme velocity.
A recent study puts complex C at $\sim7\!-\!8\kpc$ above the mid-plane \citep{Thom2008}. 
Interestingly, its projected velocity is not very different from what one would derive by extrapolating our vertical rotational gradient at these heights.

\subsection{Comparison with external galaxies}

The properties of the Galactic \hi\ halo, derived here, are similar to those of external galaxies.
The vertical rotational gradient estimated in Section \ref{application} is the same found by \citet{Oosterloo2007} for NGC\,891; inward radial motions of $10-20\kms$ have been found by \citet{Fraternali2002} in the \hi\ halo of NGC\,2403; 
\citet{Boomsma2008} showed the presence of vertical motions for halo gas clouds in NGC\,6946.
The halo \hi\ mass derived for the Milky Way is also consistent with the values estimated in other galaxies \citep[see][] {Fraternali2010}.
However, we notice that the techniques used to separate halo gas from disk gas are very different and a direct comparison is difficult.
Nevertheless, it is clear that a remarkably massive halo like that detected in NGC\,891 ($1.2\times10^9M_{\odot}$, $30\%$ of the total \hi) is not present in the Milky Way and it is perhaps fair to say at this point that NGC\,891 constitutes an extreme case.

A recurring question is whether or not the anomalous \hi\ clouds observed around external galaxies are of the same nature as the galactic IVCs and HVCs.
If all halos are made up of clouds with properties similar to the IVCs then in external galaxies they would be unresolved and appear like a smooth medium, as is observed.
Observations of external galaxies also show the presence of very massive clouds ($\sim10^7M_{\odot}$) such as the filaments of NGC\,891 \citep{Oosterloo2007} and NGC\,2403 \citep{Fraternali2002}.
These clouds have similar properties to large Galactic HVCs like complex C.
They have the same masses \citep[the \hi\ mass of complex C is $\sim5\times10^6M_{\odot}$,][]{Thom2008} and they show comparable kinematics.
For instance the filament in NGC\,2403 follows the general kinematic pattern of the halo gas but at a more extreme velocity.
This is analogous to what has been described in Section \ref{HVCsIVCs} for complex C.
In external galaxies, clouds are also detected at large distances and/or very anomalous velocities: e.g.\ the `forbidden' clouds in NGC\,2403 \citep{Fraternali2002}, the so-called counter-rotating clouds \citep[see][] {Fraternali2010} and the clouds surrounding M\,31 at tens of kpc from the disk \citep{Thilker2004}.
Very massive and very anomalous clouds are not consistent with a galactic fountain origin.
They are more likely to be gas falling into these galaxies for the first time either from satellites or directly from the IGM, although the latter mechanism is not clearly understood \citep{Binney2009}.
All things considered, they are similar to the Galactic HVCs.

\section{Conclusions}\label{conclusion}

Our Galaxy offers the unique opportunity to study an \hi\ halo from two different perspectives: on the one hand one may analyse the single bright anomalous clouds - the HVCs and IVCs - which populate the extra-planar regions, on the other hand one may consider the halo as a global component and analyse its global parameters. 
In this paper we adopted the latter approach.

We modelled the extra-planar \hi\ of the Milky Way and compared its emission with the Leiden-Argentine-Bonn Survey. Using the tangent point method we estimated the vertical gradient in the rotation velocity. The main results can be summarized as follows:
\begin{enumerate}
\item the Milky Way has a halo of neutral gas with an \hi\ mass of $3.2^{+1.0}_{-0.9}\times10^8 M_{\odot}$ ($\sim5\!-\!10\%$ of the total \hi\ mass) and an average vertical scale-height of $1.6^{+0.6}_{-0.4}\kpc$, which seems to be decreasing in the innermost regions.
\item The \hi\ halo has a global vertical lag in rotation velocity. We measured the magnitude of this lag in the region $3.5\!<\!R\!<\!5.5\kpc$ and $1\!<\!|z|\!<\!4\kpc$, obtaining an average value of $-15\pm4\kmskpc$. 
\item We detected a global inflow motion in both the vertical ($-20^{+5}_{-7}\kms$) and the radial ($-30^{+7}_{-5}\kms$) directions. 
\item The \hi\ halo is likely to have been generated by supernova explosions, which create a galactic fountain.
If a single fountain cloud has properties similar to the local IVCs, then the halo should contain thousands of these objects.
\item The properties of the Milky Way halo are broadly similar to those derived for external galaxies.
\end{enumerate}
\begin{acknowledgements} 
We thank Renzo Sancisi and Federico Lelli for helpful discussions and comments. We are grateful to an anonymous referee for his/her valuable suggestions. 
\end{acknowledgements}

\end{document}